\documentclass[%
secnumarabic,graphics,nofootinbib,tightenlines,nobibnotes,floatfix,aps,
superscriptaddress,showkeys,preprintnumbers,amsmath,amssymb,preprint,
showpacs,preprintnumbers,]{revtex4-1}

\usepackage[english]{babel}
\usepackage[dvips]{graphicx}
\usepackage[sort&compress]{natbib}
\usepackage{graphicx}
\usepackage{amsfonts}
\usepackage{amsmath}
\usepackage{amssymb}
\usepackage{amsbsy} 
\usepackage{amsmath}
\usepackage{graphicx}
\usepackage{verbatim}
\usepackage{color}
\usepackage{subfigure}
\usepackage{hyperref}
\usepackage{booktabs,caption}
\usepackage{float}
\usepackage{soul}
\usepackage[all]{xy}
\usepackage{tikz-cd}
\usepackage{color}
\usepackage{lineno}
\usepackage{hyperref}
\usepackage{amsmath}
\usepackage{amssymb}
\usepackage{slashed}
\usepackage{rotating}
\usepackage{dcolumn}
\usepackage{bm}
\usepackage{textcomp}

\def\be{\begin{equation}}
\def\ee{\end{equation}}
\def\bea{\begin{eqnarray}}
\def\eea{\end{eqnarray}}
\renewcommand{\d}{\mathrm{d}}
\newcommand{\te}{\tilde{E}}
\newcommand{\ti}{\tilde{I}}

\begin{document}

\title{Statistical origin of Legendre invariant metrics}

\author{V. Pineda-Reyes}
\email{viridiana.pineda@correo.nucleares.unam.mx} 
\affiliation{Instituto de
	Ciencias Nucleares, Universidad Nacional Aut\'onoma de
	M\'exico,\\AP  70-543, Ciudad de M\'exico, 04510 Mexico}

\author{L. F. Escamilla-Herrera}
\email{lenin.escamilla@correo.nucleares.unam.mx}
\affiliation{Instituto de
	Ciencias Nucleares, Universidad Nacional Aut\'onoma de
	M\'exico,\\AP  70-543, Ciudad de M\'exico, 04510 Mexico}

\author{C. Gruber}
\email{christine.gruber@uni-oldenburg.de} 
\affiliation{Hanse-Wissenschaftskolleg Delmenhorst, Germany}
\affiliation{Institut f\"ur Physik, Universit\"at Oldenburg, 
	D-26111 Oldenburg, Germany}

\author{F. Nettel}
\email{fnettel@ciencias.unam.mx}
\affiliation{Departamento de F\'\i sica, Fac. de Ciencias \\
	Universidad Nacional Aut\'onoma de
	M\'exico,\\AP  70-543, Ciudad de M\'exico, 04510 Mexico}

\author{H. Quevedo}
\email{quevedo@nucleares.unam.mx} 
\affiliation{Instituto de
	Ciencias Nucleares, Universidad Nacional Aut\'onoma de
	M\'exico,\\AP  70-543, Ciudad de M\'exico, 04510 Mexico}
\affiliation{Dipartimento di Fisica and ICRA, Universit\`a di Roma 
	``La Sapienza", I-00185 Roma, Italy}
\affiliation{Institute of Experimental and Theoretical Physics, 
	Al-Farabi Kazakh National University, Almaty, Kazakhstan}

\date{\today}

\begin{abstract}
Legendre invariant metrics have been introduced in Geometrothermodynamics to take into account the important fact that the thermodynamic properties of physical systems do not depend on the choice of thermodynamic potential from a \emph{geometric} perspective. In this work, we show that these metrics also have a statistical origin which can be expressed in terms of the average and variance of the differential of the microscopic entropy. To show this, we use a particular reparametrization of the coordinates of the corresponding thermodynamic phase space. 
\end{abstract}
\maketitle

\section{Introduction}
\label{sec:int}
Thermodynamics is one of the most general and successful branches of physics, in the sense that its applicability covers a wide range of physical systems with different properties and structure. However, it is a purely phenomenological theory, since it is essentially based on a set of empirical laws which are the result of far-reaching observational and experimental data. It is widely accepted that, within its range of applicability, thermodynamics well describes any macroscopic system. Despite this success, alternative approaches to thermodynamics have been constructed in the past decades, which also have proven to be useful. In particular, different geometric representations of thermodynamics have been explored \cite{1948Gibb,1975Weina,1979Rupp,2007Quev}. 
These formulations are constructed by promoting the thermodynamic space of equilibrium states, ${\cal E}$, to a differential manifold endowed with a metric structure, usually of Riemannian type. 

Weinhold \cite{1975Weina} and, a few years later, Ruppeiner \cite{1979Rupp} proposed to endow the space of equilibrium states $\cal{E}$ with metrics whose components are defined via the Hessian matrices of a given thermodynamic potential or fundamental thermodynamic relation which describes the thermodynamic system under consideration. If the potential is the internal energy $U$, the metric obtained is called the Weinhold metric $g^W$ \cite{1975Weina}; if the selected thermodynamic potential is (minus) the entropy $S$, the corresponding associated metric is called the Ruppeiner metric $g^R$ \cite{1979Rupp}. These metrics can be written as 
\be \label{hessian}  
	g^W = \frac{\partial^2 U}{\partial E^a\partial E^b} dE^a 
	\otimes dE^b \, \quad \text{and} \quad g^R = - \frac{\partial^2 S}
	{\partial E^a\partial E^b} dE^a \otimes dE^b \,.
\ee
In this work, we use the convention that repeated indices indicate a sum over all their values. The $E^a$ ($a=1,...,n$) represent the set of extensive variables necessary to describe a system with $n$ thermodynamic degrees of freedom in either representation. It can be shown that $g^R$ is conformally related to $g^W$ via the inverse of the temperature as the conformal factor \cite{1984Salamon}. 

The thermodynamic fundamental relation as a function of the extensive variables, either in the entropic or energetic representation, is completely sufficient to describe any given thermodynamic system. Any other thermodynamic potential can be obtained from the internal energy or entropy via a Legendre transformation, and thus a metric based on the Hessian of any thermodynamic potential is a suitable structure for the equilibrium space manifold \cite{2010Liu}. Another important property of equilibrium thermodynamics is thus Legendre invariance, i.e., the well-known fact that the physical properties of any thermodynamic system do not depend on the selected thermodynamic potential \cite{Callen}. It is thus self-evident to incorporate this feature into the metric structure as well, and the geometric formulation which is constructed around this invariance is known as Geometrothermodynamics (GTD) \cite{2007Quev}. 

GTD considers the invariance under Legendre transformations as the main condition to be satisfied by all the geometric structures in this formalism. The metric of the space of equilibrium states emerges as a consequence of imposing this invariance. This is achieved by representing Legendre transformations as coordinate transformations \cite{1980Arnold} and obtaining the most general solution for a metric invariant under these transformations. The $n$ extensive variables $E^a$, the $n$ intensive variables $I_a$ and the entropy $S$, as the thermodynamic potential, represent the set of all thermodynamic variables, which is denoted by $z^A=(S, E^a, I_a)$.
Then, a Legendre transformation can be represented as a coordinate transformation $z^A\rightarrow z'^{A}=(\phi', E'^a, I _a')$ defined by 
\be \label{legendretransf}
	S  = \phi' -  E'^k I'_k \,,\quad
	E^i = -  I'_i \,, E^j =  I'_j, \quad   
	I_i  = E'^i \,, I_j  = I'_j \,,
\ee
where $i \in I$, $j \in J$, $k= 1,...,i$, and $I\cup J$ is any disjoint decomposition of the set of indices $\{1,...,n\}$. In particular, for $I=\{1,...,n\}$ or $I=\emptyset$ we obtain the total Legendre transformation and the identity, respectively. It is easy to show that this transformation is well-defined \cite{2007Quev} in the sense that it leads to a non-vanishing Jacobian with $|\partial z^A /\partial \tilde z ^{A}|=1$. 

In the GTD formalism, the thermodynamic phase space (TPS) is denoted by $\mathcal{T}$, has $(2n+1)$ dimensions and coordinates $z^A$, and  is equipped with a Riemannian structure via the introduction of the metric $G=G_{AB} \d z ^A \d z^B$. We call a metric Legendre invariant if the functional form of $G$ is not modified by a Legendre transformation. This condition leads to a set of algebraic equations relating the components of the original metric $G_{AB}$ with those of the Legendre transformed metric $G' _{AB}$ \cite{2007Quev}. Solving this set of algebraic equations allows us to find several possible solutions, and one of them is, besides $G^{I}$ and $G^{II}$ established for other purposes, the metric 
\be \label{GIII}
	G^{III} = (\d S - I_a \d E^a)^2+(E^a I_a)^{2K+1} 
	\d E^a \d I_a\ \,,
\ee
where $K$ is an integer. The above metric is invariant with respect to partial and total Legendre transformations. It should be mentioned that it is not at all trivial to find Legendre invariant metrics. Since Legendre transformations do not form a group, it is not possible to apply the standard Killing approach to derive invariant geometric objects. The only possible path to follow is to solve the aforementioned set of algebraic equations, which leads to the most general Legendre invariant metric found so far as given by Eq.\,\eqref{GIII}. 

Additionally, the thermodynamic phase space ${\cal T}$ can be canonically endowed with a contact structure by means of the one-form 
\be \label{eq:eta}
	\Theta = \d S - I_a \d E^a \,,
\ee
satisfying the condition $\Theta \wedge (d\Theta)^n \neq 0$, which distinguishes $\Theta$ as a contact one-form on $\cal T$. According to Darboux' theorem \cite{1882Darboux}, this contact one-form is defined modulo a conformal factor $\Theta\rightarrow f\Theta$, where $f$ is a non-vanishing function $f:{\cal T} \rightarrow \mathbb{R}$. The contact one-form $\Theta$ is also Legendre invariant, since under a Legendre transformation it preserves the same functional form. Consequently, the phase space is defined in GTD by means of the triad $({\cal T}, \Theta, G^{III})$, which is a Riemannian contact manifold with the important property of being Legendre invariant. 

The geometry of the space $\mathcal{T}$ can be related in a consistent way with the equilibrium space $\mathcal{E} \subset \cal T$ via the introduction of the smooth embedding map, $\varphi: {\cal E}\rightarrow {\cal T}$ such that $\varphi ^*(\Theta) = 0$, where the asterisk denotes the pullback of the map. This means that ${\cal E}$ is a maximal Legendre sub-manifold, on which the relationship 
\be \label{eq:firstlaw}
	\d S = I_a \d E^a
\ee 
holds, which corresponds exactly to the first law of thermodynamics where the equilibrium condition
\be \label{eq:equil}
	I_a =\frac{\partial S}{\partial E^a} 
\ee 
is satisfied. Moreover, the Riemannian structure on $\mathcal{T}$ also induces a canonical Riemannian structure in the equilibrium space $\mathcal{E}$ by means of $\varphi^*(G^{III}) = g^{III}$. This pullback is explicitly written as
\be \label{gGTD}
	g^{III} =  \left(E^a \frac{\partial S}{\partial E^a} 
	\right)^{2K+1} \frac{\partial^2 S}{\partial E^a \partial E^b} 
	\, \d E^a \otimes \d E^b \,.
\ee 
The above metric is in general different from the Hessian metrics $g^W$ and $g^R$, presented in Eq.\,\eqref{hessian}, and the additional terms in $g^{III}$  guarantee the invariance under Legendre transformations as defined on the phase space $\mathcal{T}$. 

As shown in \cite{1990Mrug}, the geometric structure determined by the Hessian metrics \eqref{hessian} can be understood as emerging from a statistical approach associated with the family of Boltzmann-Gibbs (BG) canonical probability distributions $\rho$. 
In the same work, the geometric structures of the TPS are shown to be related to the descriptions of equilibrium and fluctuations on the manifold of equilibrium states, and the physical consequences of the microscopic statistical approach for the Ruppeiner metric \cite{1995Rupp} are explored.

In a previous work \cite{reparametrizations}, we investigated the reparametrizations of the thermodynamic variables and proposed a geometric description in the TPS. It was also found that the reparametrizations of the thermodynamic variables can be described geometrically  in the TPS through a two-rank tensor, whose pullback to the space of equilibrium states gives a metric, which in principle is related to  thermodynamic fluctuations. In this work, based on the results of \cite{reparametrizations}, we use the aforementioned two-rank tensor to find the general form of Legendre invariant metrics, establishing in this way the statistical origin of the metric \eqref{gGTD} in the formalism of GTD.

This paper is organized as follows. In Sec. \ref{sec:GenDistFunc}, a brief discussion of the most relevant results found in \cite{reparametrizations} are presented. It is shown how, via generalized reparametizations of the thermodynamic variables, the Riemannian structure of the TPS is modified. In Sec. \ref{sec:genGTD}, we discuss how these changes on the TPS lead to a metric structure, which can be connected with Legendre invariant metrics and, in particular, with the metrics found in GTD \cite{2007Quev}. Sec. \ref{sec:conclusions} is devoted to presenting the conclusions and perspectives of this work.

\section{Generalized probabilistic distribution function and Riemannian contact manifold}
\label{sec:GenDistFunc}

Reparametrizations of the extensive and intensive thermodynamic variables can be geometrically described in the TPS \cite{reparametrizations}. In fact, they are represented by distinct TPS's which have different contact and Riemannian structures, while leaving invariant the geometric structure of the space of equilibrium states. Indeed, as mentioned above, the TPS is a ($2n+1$)-dimensional contact manifold characterized by the Gibbs (contact) one-form, which in coordinates $S$, $E^a$ and $I_a$ takes the canonical form \eqref{eq:eta}. Moreover, the space of equilibrium states $\mathcal{E}$ is defined by  the  smooth embedding map $\varphi:\mathcal{E} \to \mathcal T$, along with conditions \eqref{eq:firstlaw} and \eqref{eq:equil}. Alternatively, in the TPS one can apply a total Legendre transformation, defined as in Eq.\,\eqref{legendretransf}, which transforms the Gibbs one-form into 
	\be \label{gibbsM}
	\Theta = \d \phi - E^a \d I_a,
	\ee
and the embedding conditions $\varphi^*(\Theta) = 0$ become
	\be \label{embedding1}
	\d \phi = E^a \d I_a, \quad E^a = \frac{\partial \phi}{\partial I_a}.
	\ee
It is only in $\mathcal{E}$ that the variables $\phi$, $E^a$ and $I_a$ acquire a thermodynamic meaning, namely the total Massieu potential (the total Legendre transform of the entropy potential) and the extensive and intensive variables, respectively. Therefore, the expressions in \eqref{embedding1} correspond to the first law and the equations of state for a thermodynamic system described by the fundamental relation $\phi(I_a)$, where $\phi$ is the total Legendre transform of the entropy. Regarding the Riemannian structure, a metric can be defined for the TPS as 
\begin{equation} \label{Gfr}
	G = \Theta \otimes \Theta + t \,,
\end{equation}
where in local coordinates $\Theta$ is given by \eqref{gibbsM} and $t$ is a tensor of rank two taking the form 
\begin{equation}  \label{t}
	t = \d I_a \overset{s}{\otimes} \d E^a \,,
\end{equation}
where  the product $\overset{s}{\otimes}$ is defined as
\begin{equation} \label{sotimes}
	\d x \overset{s}{\otimes} \d y \equiv \frac{1}{2} \left( 
	\d x \otimes \d y + \d y \otimes \d x \right) \,.
\end{equation}
The pullback of $G$ induces a metric $g = \varphi^*(G)$ on the space of equilibrium states $\mathcal{E}$, which in local coordinates reduces to the Hessian of $\phi$, 
	\begin{equation}  \label{gi}
	g = \varphi^*(G) = \frac{\partial^2 \phi}{\partial I_a \partial I_b}\ \d I_a \otimes \d I_b.
	\end{equation} 
In \cite{1990Mrug}, it was shown that under a total Legendre transformation, the induced metric $g$ on $\mathcal{E}$ coincides with Ruppeiner's metric, i.e., a metric whose components can be expressed in local coordinates as the negative Hessian of the entropy. 

The statistical origin of the Gibbs one-form and the metric $G$ is tied to the BG  distribution $\rho = \exp(-\phi + I_a H^a)$, where $\{H^a\}$ is a set of $n$ stochastic functions on the mechanical phase space $\Gamma$, e.g., the Hamiltonian of a system of particles. In the expression for the BG entropy, 
\begin{equation} \label{boltzmann}
	S = -\int \rho \ln \rho\  \d \Gamma \,
\end{equation}
the microscopic entropy defined as 
\begin{equation}
	s = - \ln \rho = \phi - I_a H^a
\end{equation} 
can be introduced. If the normalization of the distribution is not imposed, the microscopic entropy can be understood as a function of the variables $\{\phi, I_a\}$. If, in addition, we consider the variables $\{\phi, E^a, I_a\}$ as a set of independent variables for a higher-dimensional space, we can relate the above geometric objects to statistical quantities as follows. The Gibbs one-form is related to the statistical average of the differential of the microscopic entropy,
\begin{equation} \label{ds}
	\langle \d s \rangle \equiv \Theta = \d \phi - E^a \d I_a \,,
\end{equation}
while the metric $G$ is related to its variance
\begin{equation} \label{vards}
	\langle (\d s - \langle \d s \rangle )^2 \rangle \equiv t = 
	\d E^a \overset{s}{\otimes} \d I_a \,.
\end{equation}
Therefore, the Gibbs one-form geometrically encodes the information about equilibrium, while the metric $G$ is associated with the statistical fluctuations of the system. For a detailed description of the connection between geometry in the TPS and statistical mechanics we refer the reader to Refs. \cite{1990Mrug, HamThermo}.

Turning our attention to the description of reparametrizations in $\mathcal{E}$, in \cite{reparametrizations} it was shown that different geometric structures on the TPS can be used to determine the effects of such reparametrizations. In particular, a new contact one-form $\tilde{\Theta}$ can be chosen that is not related by a contactomorphism to \eqref{gibbsM} and a new tensor $\tilde{t}$ can be defined to account for the reparametrizations at the level of the Riemannian structure. In spite of having a different TPS, when the pullback of these new structures to the space of equilibrium states is done, the Riemannian manifold $(\mathcal{E},g)$ turns out to remain unchanged. The new geometric structures in the TPS comprising the information about the reparametrization can be expressed as follows. The new Gibbs one-form in terms of the new, tilded variables is
	\be \label{Thetat}
	\tilde{\Theta} = \d \tilde{\phi} - \te^a \d \ti_a,
	\ee
and the new metric is
	\be \label{Gt}
	\tilde{G} = \tilde{\Theta} \otimes \tilde{\Theta} + \tilde{t} \, ,
	\ee
where in the local coordinates the tensor $\tilde{t}$ takes the form
	\be \label{ttilde}
	\tilde{t} = \d \te^a \overset{s}{\otimes} \d \ti_a.
	\ee
Restricting the reparametrizations to the form $\tilde{\phi} = \phi$, $\ti_a = \ti_a(I_a)$ and $\te^a = \te^a(E^a)$, implying that the Jacobian of the transformation is diagonal, the new Gibbs one-form can be expressed in terms of the original variables as
\be \label{Thetanew}
	\tilde{\Theta} = \langle \d \tilde{s} \rangle_{\ti} = \d \phi 
	- \te^a(E^a) \frac{d \tilde I_a}{d I_a} \d I_a \,,
\ee
where $\tilde{s} = \phi - \ti_a H^a$ is the microscopic entropy in terms of the new variables. The tensor $\tilde{t}$, which geometrically encodes the information about the statistical fluctuations and the reparametrizations in $\mathcal T$, can be expressed in terms of the original variables as 
\be \label{tensort}
	\tilde{t} = \frac{d \tilde I_a}{d I_a} \ \d \te^a\ \overset{s}{\otimes} \d I_a 
	= \frac{d \tilde I_a}{d I_a} \frac{d \tilde E_a}{d E_a} \ \d E^a\ \overset{s}{\otimes} \d I_a \,.
\ee 
Therefore, we have two different contact structures \eqref{gibbsM} and \eqref{Thetanew}, as well as two different Riemannian structures \eqref{Gfr} and \eqref{Gt} on the TPS. The important feature is that these different structures have no effect on the geometric description in the space of equilibrium states -- when projected to $\mathcal{E}$ they result in the same Legendre submanifold. 

In this work, we are interested in the role that the tensor $\tilde{t}$ might play in the geometric description of statistical fluctuations. To this end, we propose to endow the TPS with a different Riemannian structure constructed as $(\mathcal{T}, \Theta, \mathcal{G})$, where the metric $\mathcal{G}$ is defined as
	\be \label{Gprime}
	\mathcal{G} = \Theta \otimes \Theta + \tilde{t} \,,
	\ee
which in the local coordinates $\{\phi, E^a, I_a\}$ takes the form
\be \label{calGcoord}
	\mathcal{G} = (\d \phi - E^a \d I_a) \otimes (\d \phi - E^b \d I_b) + \frac{d \tilde I_a}{d I_a} \ \frac{d \tilde E_a}{d E_a} \ \d E^a \overset{s}{\otimes} \d I_a \,.
\ee
This Riemannian structure, when projected to the space of equilibrium states $\mathfrak{g} = \varphi^*(\mathcal{G})$, will furnish $\mathcal{E}$ with a different Riemannian structure, in particular, different to the Hessian family of metrics widely used in the geometry of thermodynamics. It is worth emphasizing that, as we are using the Gibbs one-form \eqref{gibbsM}, the first law of thermodynamics will be expressed in its regular form in terms of the canonical variables $\{E^a, I_a\}$, i.e., as given by \eqref{embedding1}. However, the information about the fluctuations contained in the metrics $\mathcal{G}$ and $\mathfrak{g}$ differs from the information in the original Riemannian structure on $\mathcal T$.

\section{Generalized Geometry of Thermodynamics}
\label{sec:genGTD}
In the context of thermodynamics and statistical mechanics, the parameters $\{I_a\}$ of the BG probability density are identified with intensive thermodynamic variables and $\phi$ with the total Massieu potential (the total Legendre transformation of the entropy), whereas the metric \eqref{gi} gives a notion of thermodynamic length for curves connecting different states and relates the extent of the fluctuations with the geometric distance \cite{crooks}. Therefore, it is reasonable to look for a similar interpretation for the induced metric on $\mathcal{E}$ obtained from \eqref{Gprime}. 

As we mentioned in the previous section, the tensor field $\tilde{t}$
and consequently $\mathcal{G}$, captures the geometrical information of a generalized probability distribution. The objective is to translate this information into the geometric description of thermodynamic systems, i.e., to determine how this \emph{statistical} information is inherited by the geometrical structures on the submanifold of \emph{thermodynamic} states. 

In the submanifold $\mathcal{E}$, determined by the condition $\varphi^*(\Theta) = 0$, each point represents an equilibrium state for the system described by the thermodynamic fundamental relation $\phi = \phi(I_a)$, in which the first law of thermodynamics \eqref{embedding1} is satisfied. A fundamental feature of thermodynamics is its invariance under Legendre transformations. In a geometric language, this property is enclosed in the fact that the one-form \eqref{gibbsM} is invariant under such transformations on $\mathcal{T}$. 

As shown in \cite{1990Mrug}, the second law of thermodynamics is satisfied by the submanifold $\cal E$ by construction, since it is derived directly from a maximum entropy principle. In our particular case, we guarantee the second law of thermodynamics to be satisfied since $\phi(I_a)$ is the total Massieu potential. This potential is a convex function of the intensive variables $I_a$, or equivalently, its Legendre transform, the entropy, is a concave function of the extensive variables $E^a$, except at the points of macroscopic phase transitions. However, in the geometric description, the contact structure does not guarantee the second law and it is necessary to endow the contact manifold with a Riemannian structure. In the canonical case, i.e., using the family of BG probability distributions, the metric induced on $\cal E$ by the metric \eqref{Gfr} is given by Eq.\,\eqref{gi}, 
which under a total Legendre transformation coincides with the entropic metric $g^R$ of the geometric thermodynamic formulation of Ruppeiner \cite{1979Rupp,1995Rupp}, defined as the negative Hessian of the entropy as given by Eq.\,\eqref{hessian}. The metric \eqref{gi} allows us to relate the local thermodynamic stability conditions for the corresponding thermodynamic potential to a geometric structure. Moreover, as the metric is also a measure of distance between points in the manifold, the length of curves in $\cal E$ is related to the number of fluctuations along the path followed during a quasi-static thermodynamic process, larger fluctuations corresponding to a smaller distance between points \cite{crooks}.

\subsection{Generalized metrics for thermodynamics}
\label{subsec:metricG}
We recall that the generalized metric \eqref{Gprime} on the Riemannian contact manifold $(\mathcal{T},\Theta, \mathcal{G})$ is defined by means of the Gibbs form $\Theta$ and the tensor field $\tilde{t}$, whose expressions in local coordinates are given by \eqref{gibbsM} and \eqref{ttilde}, respectively. In terms of local coordinates, $\mathcal{G}$ takes the form given in \eqref{calGcoord}. Then, the induced metric $\mathfrak{g} = \varphi^*(\mathcal{G})$ on $\mathcal{E}$ is  
\be \label{gcalmetric}
	\mathfrak{g} = \varphi^*(\mathcal{G}) = \frac{d \tilde I_a(I_a)}{d I_a}
	\frac{d \tilde E^a(E^a)}{d E^a}{}\bigg|_{E^a = \partial \phi/\partial I_a}
	 \ \frac{\partial^2 \phi}{\partial I_a \partial I_b}\ \d I_a \overset{s}
	 {\otimes} \d I_b \,,
\ee 
Equations \eqref{Gprime} and \eqref{gcalmetric} are in fact a family of metrics for $\mathcal{T}$ and $\mathcal{E}$, respectively. Indeed, for each set $\{\ti_a = \ti_a(I_a), \te^a = \te^a(E^a)\}$ we have a different metric structure on $\mathcal{T}$ as well as on $\mathcal{E}$. Thus, \eqref{gcalmetric} represents a family of thermodynamic metrics on $\mathcal{E}$ and each one is related to the statistical fluctuations of the system and contains information about the specific reparametrization.

\subsection{Legendre invariant metrics}
\label{subsec:g3}
The invariance of the contact structure under a Legendre transformation incorporates the well-known thermodynamic fact that the thermodynamic information of any physical system described by a fundamental relation is preserved under any such  transformations. In geometric terms, this corresponds to the invariance of the Gibbs one-form in the TPS under such a transformation. That is, the equilibrium description is Legendre invariant by construction. On the contrary, the metric \eqref{Gfr} is not invariant on $\mathcal{T}$, in the sense that it does not preserve its form under the action of a Legendre transformation, a property that the metric \eqref{gi} inherits. Therefore, we conclude that the geometric description of fluctuations on $\mathcal{E}$ is not invariant under Legendre transformations. In fact, it is known that Hessian metrics in the geometry of thermodynamics obtained from different (Legendre related) potentials correspond to different behaviors in the various ensembles \cite{2010Liu, termometrica}. 

The behavior of the family of metrics \eqref{Gprime} is now analyzed with respect to a general Legendre transformation, in order to determine the requirements for $\mathcal{G}$ to remain invariant under such transformations. This family of metrics has a freedom in the choice of the functions $\{\ti_a = \ti_a(I_a), \te^a = \te^a(E^a)\}$ and unlike \eqref{Gfr}, there is the possibility of finding an invariant metric under Legrendre transformations. We can say that this requirement amounts to asking for a invariant geometric description of fluctuations in terms of Riemannian structures on the manifolds $\mathcal{T}$ and $\mathcal{E}$. 

A partial Legendre transformation for the set of coordinates $\{\phi,I_a,E^a\}$ is given by Eq.\,\eqref{legendretransf},
where the same conventions for the set of indices $\{i,j\}$ are applied, and $\phi'$ becomes the new thermodynamic potential on $\mathcal{E}$. The generalized metric $\mathcal{G}$ transforms as
\be \label{Gtildetrans}
	\mathcal{G} = \Theta \otimes \Theta + \frac{d \tilde I_i(E'{}^i)}{d E'{}^i} 
	\frac{d \tilde E ^i(-I'{}_i)}{d I'{}_i} \d E'{}^i \overset{s}{\otimes}
	\d I'{}_i +\frac{d \tilde I_j(I'{}_j)}{d I'{}_j} \frac{d \tilde
	E^j(E'{}^j)}{d E'{}^j} \d E'{}^j \overset{s}{\otimes} \d I'{}_j \,,
\ee
with $\Theta = \d \phi' - I'{}_a \d E'{}^a$. It is straightforward to show that in general this metric is not covariant under a general Legendre transformation. The fulfillment of  this property will clearly depend on the functions $\tilde I_a(I_a)$ and $\tilde E^a(E^a)$. In order to obtain a form of $\mathcal{G}$ which is covariant under the full set of Legendre transformations, two conditions must be satisfied by the set of functions $\{\tilde I_a, \tilde E^a\}$: (i) all of $\tilde I_a$ and $\tilde E^a$ must have the same functional form, and (ii) $\tilde I_a$ and $\tilde E^a$ must be even functions, i.e., $\tilde E^a(-E^a) = \tilde E^a(E^a)$ and $\tilde I_a(-I_a) = \tilde I_a(I_a)$. A particular choice for the set of functions satisfying the above conditions is 
\be \label{choicegf}
	\tilde I_a(I_a) = \frac{(I_a)^{2K +2}}{2K+2}, \quad \text{and} 
	\quad \tilde E^a(E^a) = \frac{(E^a)^{2K+2}}{2K+2}\,,
\ee
with $K$ an integer. This choice for $\tilde I_a$ and $\tilde E^a$ yields the following metric for the thermodynamic phase space $\cal T$,
\be \label{G3}
	\mathcal{G} = \Theta \otimes \Theta + \left(I_a E^a\right)^{2K +1} 
	\d E^a \overset{s}{\otimes} \d I_a\,.
\ee
This metric was proposed within the formalism of GTD and coined as $G^{III}$; its main feature is to be invariant under any Legendre transformation.
The corresponding induced metric on the space of thermodynamic equilibrium states $\cal E$, \emph{after a total Legendre transformation}, is then 
\begin{equation} \label{g3metric}
	\mathfrak{g} =\varphi^*(\mathcal{G})= \left( E_a \frac{\partial S} 
	{\partial E_a} \right)^{2K +1} \frac{\partial^2 S}{\partial E_a 
	\partial E_b} \ \d E_a \otimes \d E_b \,.
\end{equation}
This is the same metric as the one induced via the pullback $g^{III} = \varphi^*(G^{III})$, given in Eq.\,\eqref{gGTD}. It provides a notion of distance between thermodynamic states on the submanifold $\cal E$, which is different from the one given by the Hessian thermodynamic metrics. However, each component of the above metric is still proportional to the second derivatives of the thermodynamic potential with respect to its corresponding variables, thus providing a connection between distances in $\cal E$ and thermodynamic fluctuations, as well as between curvature singularities and phase transitions; nonetheless, these relations between geometric and thermodynamic features are not completely understood yet, nor is it clear how to establish them from the general geometric expressions. Nonetheless, there is enough evidence and a large number of results supporting these claims \cite{2015Quev,2013Brava,2011Quevb,2012Aviles,2014Quevedo,2010Vasq,2015Quevedo}.

\section{Conclusions}
\label{sec:conclusions}

To develop a geometric description of equilibrium thermodynamics two different classes of Riemannian metrics have been applied, namely, Hessian metrics and Legendre invariant metrics. The first class was created by introducing metrics into the equilibrium space, whose components coincide with the Hessian matrix of specific thermodynamic potentials. In particular, the Hessian metric derived from the entropy (the Ruppeiner metric) has been shown to have a very interesting interpretation in terms of statistical quantities derived from the BG  entropy. 

The class of Legendre invariant metrics have been introduced in the formalism of GTD with the aim of taking into account the well-known property that in equilibrium thermodynamics the physical properties of a system do not depend on the thermodynamic potential used to describe the system. This class of metrics has been introduced by using a purely geometric approach in which the thermodynamic phase space is defined as a Riemannian contact manifold. 

In this work, we have shown that Legendre invariant metrics can also be interpreted in terms of statistical quantities, namely, the average and the variance of the differential of the microscopic entropy. To this end, we introduce a particular reparametrization of the extensive and intensive variables, which are used as coordinates in the thermodynamic phase space. These reparametrizations are encoded in a two-rank tensor, which is shown to contain information about the statistical fluctuations of the system. We conclude that Legendre invariant metrics can also be interpreted in terms of statistical quantities that are related to the thermodynamic fluctuations of the system under consideration.

The approach presented here can be applied not only to Legendre invariant metrics, but also to any metric of the phase space that can be obtained by the particular parametrization, which involves only the extensive and intensive variables, separately. It would be interesting to analyze the properties of this new class of metrics and explore their applicability in the context of the  geometric descriptions of thermodynamics.

\section*{Acknowledgments}

L. F. Escamilla-Herrera and F. Nettel thank the financial support from the Consejo Nacional de Ciencia y Tecnolog\'ia (CONACyT, Mexico). C. Gruber acknowledges support by a Junior Fellowship of the Hanse-Wissenschaftskolleg Delmenhorst, and from the University of Oldenburg and the Research Training Group "Models of Gravity". This work was partially supported  by UNAM-DGAPA-PAPIIT, Grant No. 111617.

\bibliographystyle{ieeetr}
\bibliography{G3}

\end{document}